\documentclass{aa}  

\usepackage{graphicx}
\usepackage{txfonts}
\usepackage{lipsum}
\usepackage{subcaption}         % necessary for continued figures, example in section 3
\usepackage{lscape}             % to rotate a single page table, example in appendix.
\usepackage{placeins}           % useful with \FloatBarrier, to keep 
\def\lsim{\raise0.3ex\hbox{$\;<$\kern-0.75em\raise-1.1ex\hbox{$\sim\;$}}}
\def\gsim{\raise0.3ex\hbox{$\;>$\kern-0.75em\raise-1.1ex\hbox{$\sim\;$}}}

\def\be{\begin{equation}}
\def\ee{\end{equation}}

\def\theta{\vartheta}

\def\R{{\cal R}}

\def\d{{\rm d}}

\def\C{Cygnus~X-3}

\renewcommand{\vec}[1]{\boldsymbol{#1}}

\usepackage{xcolor,color}
\linenumbers

\begin{document}

   \title{Cygnus~X-3 as semi-hidden PeVatron}

   %\subtitle{}

   \author{M.~Kachelrie\ss\inst{1}
     \and  E.~Lammert\inst{1,2}}

   \institute{Institutt for fysikk, NTNU, Trondheim, Norway
            \and Department of Physics, School of Natural Sciences,
  Technical University Munich, Germany\\ }

   \date{}

% \abstract{}{}{}{}{}
% 5 {} token are mandatory
 
  \abstract
      {The high-mass X-ray binary \C\ has been suggested for a long time to be
        a source of high-energy photons and neutrinos.}
   {In view of the increased
sensitivity of current experiments, we examine the acceleration and
interactions of high-energy cosmic rays (CRs) in this binary system, assuming
that the compact object is a black hole.} 
   {        Using a test-particle approach in
a Monte-Carlo framework, we employ as the
basic CR acceleration mechanisms  magnetic reconnection or 2.nd order
Fermi acceleration  and diffusive shock acceleration. 
}
   {We find that in all three scenarios CRs can be accelerated beyond PeV energies.
High-energy photons and neutrinos are produced as secondaries in
photo-hadronic interactions of CRs on X-ray photons and in the scattering
on gas from the wind of the companion star.
Normalising the predicted photon flux to the excess flux observed by
LHAASO at energies above PeV in the direction of \C, a CR acceleration
efficiency of $10^{-3}$ is sufficient to power the required CR luminosity.
Our results suggest that the PeV photon flux from \C\  could be in a bright
phase significantly increased relative to the average flux of the last years.
}
   {}


   \maketitle

%%%%%%%%%%%%%%%%%%%%%%%%%%%%%%%%%%%%%%%%%%%%%%%%%%%%
\section{Introduction}
%%%%%%%%%%%%%%%%%%%%%%%%%%%%%%%%%%%%%%%%%%%%%%%%%%%%%
%
Cygnus~X-3, a high-mass X-ray binary in the direction of the Cygnus~OB
association, has had an outstanding impact on the development of gamma-ray
astronomy. In 1983, \citeauthor{1983ApJ...268L..17S} found a periodically
modulated flux of neutral particles with energies above $2\times 10^{15}$\,eV
coincident with Cygnus~X-3 in archival data from the Kiel air-shower array.
This apparent detection and the supporting evidence provided during the
following years by various experiments operating in the northern hemisphere
triggered an enormous theoretical interest
in this source~\citep{Gaisser:1983cj,Berezinsky:1985zja,1986ApJ...301..235B}.
More importantly, these early results on Cygnus~X-3
stimulated the building of air-shower arrays designed specifically
for gamma-ray astronomy. However, none of these newly built arrays, like
CASA-MIA and SPASE, detected a signal from  Cygnus~X-3, for a review of 
these early results see~\citet{1994ApJS...90..883P}. Moreover,  
theoretical arguments suggest that most of these claims, in
particular those by underground experiments, were erroneous
as discussed already by~\citet{Berezinsky:1985hr}.
Still, it is a tantalising option that the initial claims for a photon
signal were correct. In this case, Cygnus~X-3 might have been in a bright
phase in the early 1980s, which ended before the start of
data-taking of the following generation of experiments. Such a strong
variability is consistent with the decrease
by at least a factor of 100 reported in the long-term average flux
between the mid-1970s and mid-1980s~\citep{1986ApJ...306..587B}.

Around 40\,years later, the advent of advanced air-shower arrays with
improved hadron-photon separation power has started a new era in very-high
energy
gamma-ray astronomy. In particular, LHAASO has detected high-energy photons
with energies up to few PeV in the direction of the Cygnus~X
superbubble~\citep{LHAASO:2023uhj}. Moreover, the measured photon flux from
the central region of the bubble is enhanced, with two out of eight PeV photons
located inside a region of radius $0.5^\circ$. This excess, which corresponds
to an increased flux by a factor of 5--10 compared to the average from the
Cygnus superbubble, points towards additional point sources in this central
region. The aim of this work is to examine if, and under which conditions,
\C, which is located in the  central region of the Cygnus superbubble,
may be responsible for this excess PeV photon flux.

Using a phenomenological test-particle approach, we employ depending on the
level of magnetisation in the jet different mechanisms for the acceleration
of cosmic rays (CRs): In the strong magnetic turbulence close to the black
hole (BH), we assume that 2.nd order Fermi acceleration or magnetic
reconnection are efficient acceleration processes, while we suppose that
diffusive shock acceleration (DSA) operates in the environment with small
magnetisation at the termination shock of the two jets emanating from the BH.
We perform simulations for two states of \C, which differ among others by
the spectral energy distribution of X-ray photons.
In the first one, called $S_1$, the jet has
switched on right after the quiescent state, interacting with the
surrounding medium close to the BH. In the second one, called $S_2$
and suggested by \citet{Koljonen:2023xfn} as a promising state for the
production of high-energy secondaries,
the jet has been on for some time, working against the Wolf-Rayet medium,
forming a cocoon and a termination shock. Thus the  soft X-ray state $S_1$
is most likely disk-dominated, while the hard X-ray state $S_2$ corresponds
to a corona/jet-dominated state.
We simulate the acceleration and interactions of CRs in
a Monte-Carlo framework, including both hadronic and photo-hadronic
interactions. For the high-energy photons which are produced as secondaries
in these interaction we include absorption via pair production in the
photon fields present.

The plan of this article is as follows: We start in Section~2  with a brief
description of \C, summarizing our choice for the values of the relevant
physical parameters which enter
in the following calculations. Then we describe in Section~3 the
acceleration scenarios we assume to operate in \C, and the
Monte-Carlo scheme we employ to follow the time evolution of particles.
Finally, we present in Section~4 the resulting
time scales of the relevant acceleration, interactions and energy loss
processes, and our numerical
results for the fluxes of high-energy photons and neutrinos. We finish
by a discussion of the results and the underlying assumptions,
before we conclude.

%%%%%%%%%%%%%%%%%%%%%%%%%%%%%%%%%%%%%%%%%%%%%%%%%%%%%%%%%%%%%%%%%%%%%%%%%%%%%
\section{Environment of Cyg~X-3}             
%%%%%%%%%%%%%%%%%%%%%%%%%%%%%%%%%%%%%%%%%%%%%%%%%%%%%%%%%%%%%%%%%%%%%%%%%%%%
%

%%%%%%%%%%%%%%%%%%%%%%%%%%%%%%%%%%%%%%%%%%%%%%%%%%%%%%%%%%%%%%%%%%%%%%%%%%%
\paragraph{Binary system}%
%
%%% Masses:
%
The determination of radial velocity curves for the \C\ binary system
is complicated by strong optical extinction  and the difficult separation
of wind features from those arising in the photosphere of the companion.
Therefore, the individual
masses of the components, $M_{\rm CO}$ of the compact object and $M_{\rm WR}$
of the companion Wolf-Rayet (WR) star, have remained uncertain up-to-date.
Published results span a
wide range from a BH with mass close to or above
$20\,M_\odot$ \citep{1996A&A...311L..25S,Hjalmarsdotter:2007bx}
down to upper limits of $3.6\,M_\odot$~\citep{Stark:2003vr}. In contrast,
the mass ratio of the binary stars can be constrained more precisely
as $M_{\rm CO}/M_{\rm WR}=0.24\pm 0.06$ combining radial velocity curves
derived from  FeXXVI emission lines with infrared HeI absorption
lines~\citep{Hanson:2000rg}. 
Luckily, several parameters important for the determination of the
spectra of high-energy
particles have a rather weak dependence on the
masses of the binary system.  For instance, the orbital separation $a$, which
in turn influences the target density of stellar photons and the gas in the
stellar wind, depends only as $a\propto (M_{\rm CO}+M_{\rm WR})^{1/3}$ on
the  total mass of the binary system.
To be able to compare easier with earlier studies, we will use values close
to the upper mass range discussed in the literature, setting
$M_{\rm CO}=M_{\rm BH}=20M_\odot$ and $M_{\rm WR}=50M_\odot$
corresponding to the high-mass solution of ~\citet{Szostek:2007ke}.
We will disuses later how our results change if these masses are reduced.
Thus the Schwarzschild radius of the BH is $R_s= 6\times 10^{6}\,$cm,
the orbital separation of the stars in the binary system is
$a=3 \times 10^{11}$\,cm, while we use 9.6\,kpc as distance to
\C~\citep{Reid:2023ksq}.

%%%%%%%%%%%%%%%%%%%%%%%%%%%%%%%%%%%%%%%%%%%%%%%%%%%%%%%%%%%%%%%%%%%%%%%%%%%
\paragraph{Wind}%
The gas in the stellar wind of its WR companion provides an important
target for the production of secondary particles in
hadronic collisions. Mass conservation implies for the density profile of a
spherically symmetric wind
\be
n(r)=\frac{\dot M_{\rm wind}}{4\pi r^2 \mu m_p v_{\rm W}}
\ee
where we use %for the numerical estimates as mass loss rate
$\dot M_{\rm wind}=0.6\times 10^{-5}M_\odot$/yr  and as wind velocity
$v_{\rm W}=2\times 10^8$cm/s~\citep{Szostek:2007ke,Vilhu:2021vfs}.
For a proton rich environment, $\mu\simeq 1$, while for a helium
rich environment, as suggested first by~\citet{1992Natur.355..703V},
it is $\mu\simeq 4$.

%%%%%%%%%%%%%%%%%%%%%%%%%%%%%%%%%%%%%%%%%%%%%%%%%%%%%%%%%%%%%%%%%%%%%%%%%%%
\paragraph{Photon fields}%
Another important target for the production of secondary particles
are background photons from the stellar light of the WR star, from thermal
emission of the accretion disk of the BH and from the synchrotron corona.
These photon fields lead in photo-hadronic interactions to the production
of secondary photons and neutrinos, and provide a target for
the fragmentation of helium nuclei into nucleons via photo dissociation.
For the thermal photons from the WR star, we use a Planck distribution with
the temperature $T=10^5$\,K which we rescale as $(R_{\rm WR}/r)^2$ at the
distance $r$ to the WR star with radius
$R_{\rm WR}=6\times 10^{10}$\,cm~\citep{Sahakyan:2013opa}.

\begin{figure}
  \centering
  \includegraphics[width=0.49\columnwidth]{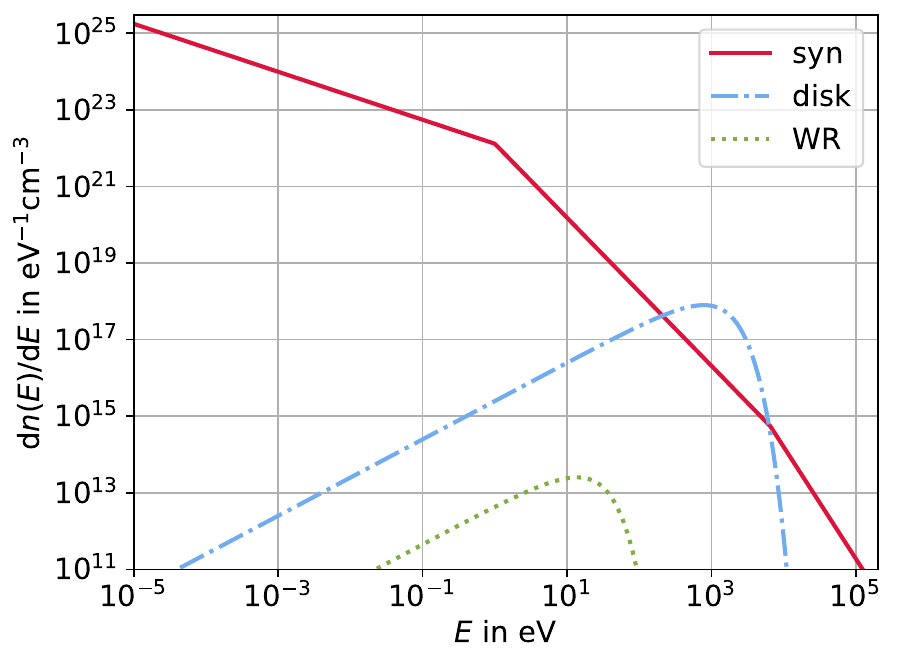}
  \includegraphics[width=0.49\columnwidth]{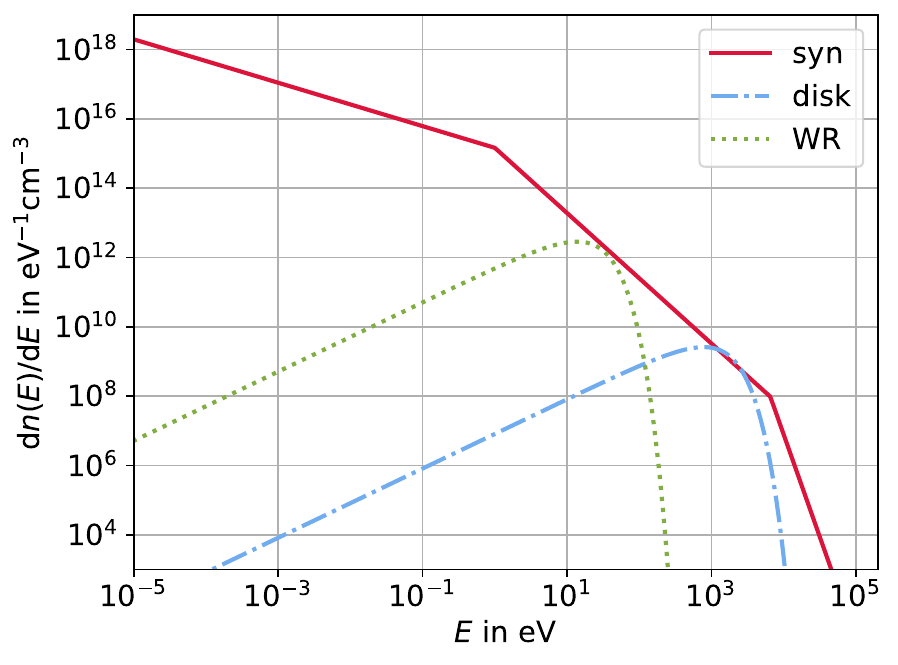}
  \caption{Spectral density of photon backgrounds in the state $S_1$ (left) and $S_2$ (right).}
  \label{fig:nbg}
\end{figure}

The accretion rate $\dot M$ in a X-ray binary may be strongly time-dependent,
implying that also the nature of the accretion disk and thus the temperature
profile of the disk varies. Since we are mainly interested in the bright
phase of \C\ with accretion close to the Eddington rate, we use a geometrically
thin, optically  thick Keplerian accretion
disk~\citep{Shakura:1972te,Chakrabarti:1996cc}. 
In this case, the thermal emission from the accretion disk is described by
the temperature profile 
\be \label{T}
 T(r) = \left( \frac{3GM\dot M}{8\sigma\pi r^3}
                \left[ 1-(R_0/r)^{1/2} \right] \right)^{1/4} .
\ee
Since the radial extension of the disk is rather small, the total
emission is close to a Planck distribution with $T(r)\simeq T(R_0)$.
For a helium-rich composition and the standard value $\eta=0.1$ for
the accretion efficiency $\eta=L_{\rm Edd}/(\dot M_{\rm Edd} c^2)$, the
Eddington luminosity becomes
$L_{\rm Edd}\simeq 2.6\times 10^{38} (M_{\rm BH}/M_\odot)$erg/s which in turn
fixes $\dot M=\lambda\dot M_{\rm Edd}$ for a chosen $\lambda$.
The observed X-ray luminosity favours a large  value of $\lambda$ and
$M_{\rm BH}$~\citep{Veledina:2023zho}.
We calculate the density of disk photons at a given point in the jet
integrating the photon emissivity into the corresponding solid angle
$\d\Omega$ over the accretion disk~\citep{2011A&A...529A.120C}, choosing
$\lambda=0.5$ and as smallest radius $R_0$ of the acceleration and of the
emission region six Schwarzschild radii,
$R_0= 6R_s\simeq 3.6\times 10^{7}\,$cm, while we set $R_{\rm out}=10R_s$
for the outer radius of the disk. 

Finally, we have to fix the spectral density of X-ray photons in the
corona. We identify the emission zone where energetic electrons are
accelerated and radiate these photons via synchrotron and inverse Compton
scattering with the acceleration region of hadrons.
Then we use the INTEGRAL measurements from~\citet{Cangemi:2020jaj}
for the quiet state $S_1$ and  for the state $S_2$ to determine
the spectral number density of X-ray photons in the range 10-100\,keV.
In order to connect them to observations in the radio range from
AMI-LA and RATAN~\citep{2012A&A...545A.110P}, 
we assume an additional break at $E=1$\,eV.
For the size of the emission region, we assume $L=8R_S$ in the
case of $S_1$ and  $L=\tan\theta R_{S_2}$ with $R_{S_2}=7\times 10^{11}$\,cm
as the distance to the termination shock and $\theta=12^\circ$ as the
jet opening angle~\citep{Mioduszewski:2001ev,10.1093/mnras/stac666}.

The resulting spectral number density of stellar, accretion disk and X-ray
photons are shown in Fig.~\ref{fig:nbg} for the state $S_1$ (left panel) and
$S_2$ (right panel).

%%%%%%%%%%%%%%%%%%%%%%%%%%%%%%%%%%%%%%%%%%%%%%%%%%%%%%%%%%%%%%%%%%%%%%%%%%%
\paragraph{Magnetic field}%
We use as radial profile for the
magnetic field strength around the BH
\be
B(r)=B_0\left(\frac{r_0}{r}\right)^\delta ,
\ee
setting $B_0=2.1\times 10^7$\,G at the jet injection point
$r_0=10R_s\simeq 6\times 10^7$\,cm~\citep{Miller-Jones:2003lop}. The slope
$\delta$ is rather uncertain, with $0.5\leq\delta\leq 0.83$ being
considered by~\citeauthor{Miller-Jones:2003lop}  Here, we
follow~\citet{Koljonen:2017gah} who argued that $\delta\simeq 0.65$
fits best data.

In addition to the field strength, the ratio between the energy density
in the magnetic field and the total energy density of the plasma is an
important parameter. This ratio, the so-called magnetisation
$\sigma_{\rm m}=B^2/(4\pi \rho)$ with $\rho\simeq n\gamma mc^2$ as the enthalpy
density, becomes of order one at $r\simeq 10^{11}$\,cm. Since the Alfv\'en
velocity $v_A=c\sqrt{\sigma_{\rm m}/(1+\sigma_{\rm m})}$ approaches the speed
of light
for $\sigma_{\rm m}\gg 1$, the borderline between weak and strong magnetisation
separates also the parameter space where DSA is more efficient than
acceleration by second-order Fermi process or magnetic reconnection.

%%%%%%%%%%%%%%%%%%%%%%%%%%%%%%%%%%%%%%%%%%%%%%%%%%%%%%%%%%%%%%%%%%%%%%%%%%%%
\section{Theoretical framework}             
%%%%%%%%%%%%%%%%%%%%%%%%%%%%%%%%%%%%%%%%%%%%%%%%%%%%%%%%%%%%%%%%%%%%%%%%%%%%
%
%%%%%%%%%%%%%%%%%%%%%%%%%%%%%%%%%%%%%%%%%%%%%%%%%%%%%%%%%%%%%%%%%%%%%%%%%%%
\subsection{Acceleration scenarios}
Various phenomenological models have been proposed to explain the
observed gamma-ray emission from \C.  %in the GeV range 
Several of them connect the gamma-ray emission to particle acceleration
in shocks,  either generated internally in the jets or externally at the
recollimination or the termination of the
jet~\citep{Romero:2003td,2012A&A...545A.110P,Baerwald:2012yd}.
If these shocks are collisionless, DSA can lead
to particle acceleration. Estimates for the velocity of the outflow in \C\
range from non-relativistic ($\beta=v/c<0.3$~\citep{1996AJ....112.2690W})
to mildly relativistic ($\beta=0.63$~\citep{Miller-Jones:2003lop} and
$\beta>0.81$~\citep{Mioduszewski:2001ev}), indicating time-dependent
jet velocities.  Such trans-relativistic
flows with ${\cal O}(\beta)\sim 1$ are well suited for fast acceleration,
if the magnetisation $\sigma_{\rm m}$ is small.
Thus we assume that DSA operates only at large
enough distance, $r\gsim 10^{11}$\,cm, from the BH, such that
$\sigma_{\rm m}\lsim 1$.
In the Bohm diffusion limit,
$D=c\R /(3B)$, the acceleration rate of a particle with
rigidity $\R$ is given by~\citep{Lagage:1983zz}
\be  \label{acc1}
 t_{\rm sh}^{-1} =  \frac{1}{E}\frac{\d E}{\d t}_{\rm sh}\simeq
 \zeta^{-1} \frac{v_{\rm sh}^2}{D}\simeq
 \frac{3B v_{\rm sh}^2}{\zeta  c\R}\simeq
 \eta \frac{cB}{\R} .
 \ee
With $\zeta\simeq 10-20$ for a parallel shock, $(v_{\rm sh}/c)^2\simeq 0.1$,
we set conservatively $\eta =10^{-3}$. Note that we assume shock velocities
$v_{\rm sh}$ small enough that relativistic effects like  beaming or
a large loss-cone of CRs at shock crossing can be neglected.

An alternative acceleration process which has recently attracted increased
attention is magnetic reconnection. This acceleration mechanism has been
applied to micro-quasars in general by \citet{deGouveiaDalPino:2003mu}
and to \C\ by \citet{Khiali:2014joa}. In this model, magnetic reconnection
happens in current sheets produced when field lines arising from the
accretion disk and of the BH magnetosphere encounter. If the
reconnection velocity is fast enough, trapped particles between
the two converging magnetic fluxes of opposite polarity can gain energy
in a similar way to the first-order Fermi process, leading to
\citep{Kow12}
\be \label{acc2}
 t_{\rm rec}^{-1} = \frac{1}{E}\frac{\d E}{\d t}_{\rm rec}\simeq
 1.3\times 10^5 \left( \frac{\R}{m_pc} \right) ^{-0.4}
 \frac{v_A}{L_{\rm rec}}
\ee
with $v_A$ as the Alfv\'en velocity  and $L_{\rm rec}$ as
the extension of the reconnection zone.
In this scenario, the acceleration region is close to the BH, at a distance
between $\simeq 6 R_S$ and $\simeq 10 R_S$. As a result, the photon fields
as potential targets for photo-hadronic interactions are much more intense
than in the DSA case. Moreover, the magnetisation is very large and thus we
assume that DSA is not effective.

\begin{figure*}[bt]
\includegraphics[width=0.99\columnwidth,angle=0]{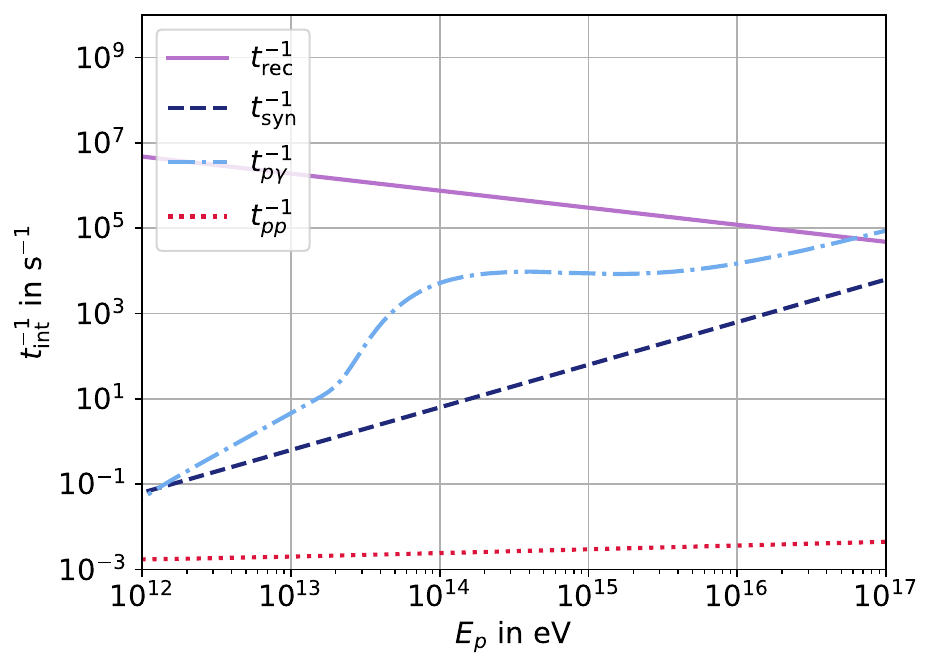}
\hskip0.5cm
\includegraphics[width=0.99\columnwidth,angle=0]{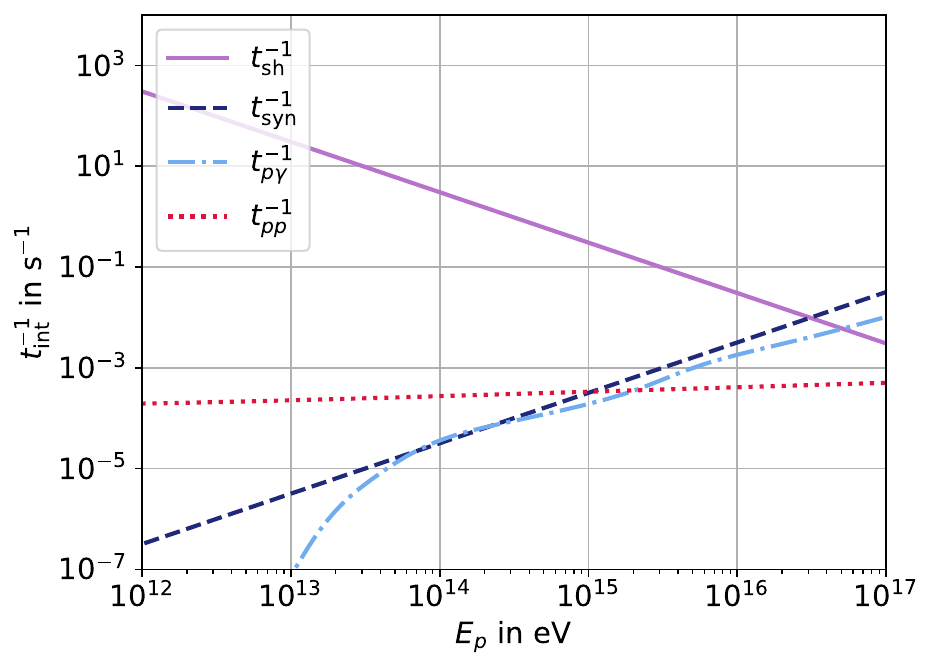}
\caption{Rates for synchrotron losses $t^{-1}_{\rm syn}$, photo-hadronic
  $t^{-1}_{p\gamma}$ and hadronic $t^{-1}_{pp}$ interactions of proton as
  function of energy compared to the acceleration rate by reconnection
  $t^{-1}_{\rm rec}$ and  2.nd order Fermi acceleration $t^{-1}_{\rm Fermi}$
  in the state $S_1$ (left) and by DSA $t^{-1}_{\rm sh}$ in the state $S_2$
  (right).}
\label{fig:rates}
\end{figure*}

Finally, we consider as alternative to magnetic reconnection  2.nd order
Fermi acceleration in the strong magnetic turbulence close to the BH.
For $\sigma_{\rm m}\gg 1$ and thus $v_A\simeq c$,
%approaches the speed of light. In this limit,
the efficiency of 2.nd order
Fermi acceleration is not suppressed relative to DSA. Moreover the slope
of the produced particle distributions can become potentially universal,
with $\alpha\sim 2.1$ found by ~\citet{Comisso:2024ymy}, and 
acceleration rate
\be \label{acc3}
 t_{\rm Fermi}^{-1} = \frac{1}{E}\frac{\d E}{\d t}_{\rm Fermi}\simeq
  4\kappa (\Gamma\beta)^2 \frac{v_A}{L_c}
 \simeq 10^{5} {\rm s}^{-1} .
\ee
Here, $\kappa$ is an efficiency factor determined by~\citet{Comisso:2019frj}
using PIC simulations as  $\kappa\sim 0.1$,
$|\vec u|=\Gamma|\vec\beta|\sim 1$ is the spatial part of the
four-velocity of the scattering centers, and $L_c$ the coherence length
of the turbulent magnetic field. We have assumed that the energy of the
magnetic turbulence is mainly contained in large-scale modes, and have
used $L_c=10^6$\,cm for the numerical estimate.
Thus these two processes lead for the chosen parameters roughly to the
same maximal energy, cf.\ with the left panel of Fig.~\ref{fig:rates},
and the same secondary production. However, second-order Fermi
acceleration has the advantage that it can operate also at
larger distances from the BH, as long as the jet is strongly magnetised.
Since at larger distances both the acceleration efficiency and the secondary
production would be reduced, the results from this acceleration mechanism
would interpolate between those of magnetic reconnection very close to the
BH and DSA at the termination shock.
Thus we consider in our simulations only these two extreme cases.

%%%%%%%%%%%%%%%%%%%%%%%%%%%%%%%%%%%%%%%%%%%%%%%%%%%
%%%%%%%%%%%%%%%%%%%%%%%%%%%%%%%%%%%%%%%%%%%%%%%%%%%

%%%%%%%%%%%%%%%%%%%%%%%%%%%%%%%%%%%%%%%%%%%%%%%%%%%%%%%%%%%%%%%%%%%%%%%%%%%
\subsection{Monte Carlo scheme}

We use a ``leaky-box scheme'' where for a chosen time step $\Delta t$ first
the escape probability of a charged particle with energy $E$ from the
acceleration region is calculated,
\be
 p_{\rm esc}= 1-\exp[ (1-\alpha)\ln(1+\xi) ] .
\ee
Here, $\alpha$ is the theoretically predicted slope of the differential
energy spectrum of accelerated particles in the case of no interactions
and energy losses, which we choose for both acceleration mechanisms as
$\alpha=2.3$, while $\xi=(\d E/\d t)_{\rm acc} \Delta t/E$ is the energy
fraction gained according to Eqs.~(\ref{acc1}) and (\ref{acc2}). If the
particle stays in the acceleration process, the true energy gain is
calculated as the sum of the acceleration gain and the continuous energy
losses, $[(\d E/\d t)_{\rm acc}-(\d E/\d t)_{\rm CEL}] \Delta t$, where the
latter include synchrotron and, for muons, inverse Compton losses. Finally,
it is decided if the particle decays or interacts: First, the
probability that something happens is compared to a random number $r$
uniformly distributed in $[0:1[$, then a second random number $r$ is
compared to the branching ratios for the relevant decay and interaction
channels.

If a photo-hadronic interactions occurs, we use the modified version of
the SOPHIA program~\citep{Mucke:1999yb} developped
by~\citet{Kachelriess:2007tr} to generate secondary particles, while
we employ QGSJET-IIc \citep{Ostapchenko:2010vb,Ostapchenko:2013pia}
for hadronic interactions. We take into account the finite decay time
of all unstable particles except neutral pions and include
only stable charged all particles in the acceleration process.
For the decay of unstable particles, we employ SIBYLL~\citep{Fletcher:1994bd}.

After the particles escaped from the acceleration region, we assume that the
magnetic field strength is small enough that they move ballistically.
Except for photons, the interaction depth is small, $\tau\ll 1$, and multiple
interactions can be neglected. Thus we add after escape a final interaction
for hadrons
with probability $p=1-\exp(-\tau)$ and let all unstable particles decay.
In the case of photons, we calculate the pair-production probability
on background photons in the direction of the observer. In addition, we add
the pair-production probability on photons of the cosmic microwave
background (CMB) and the extragalactic background light
from~\citet{Franceschini:2008tp}
to obtain the final flux observed on Earth, $F=F_0\exp(-\tau_{\gamma\gamma})$.
We neglect the absorption on the starlight in the Milky Way, which adds
only a minor correction relative to the absorption on the
CMB~\citep{Vernetto2016} and the effect of
electromagnetic cascades what is well justified above the pair-production
threshold, $E_\gamma \simeq (1-10)$\,GeV.

Note that the acceleration time is much smaller than the orbital period
of the system. Thus there is an orbital modulation of the fluxes, both because
the interaction depth varies with the orbital phase and, in addition for
photons, because 
the absorption probability changes. In the energy range we are mainly
interested in, $(10^{14}$--$10^{16})$\,eV, the modulation is however only
minor: The photon absorption probability is dominated by the CMB, while
the interaction depths vary only by a factor a few, cf.~\citet{EL25} for
details.
In the following, we present therefore results for the average fluxes.

%%%%%%%%%%%%%%%%%%%%%%%%%%%%%%%%%%%%%%%%%%%%%%%%%%%%%%%%%%%%%%%%%%%%%%%%%%%%%
\section{Numerical results}
%%%%%%%%%%%%%%%%%%%%%%%%%%%%%%%%%%%%%%%%%%%%%%%%%%%%%%%%%%%%%%%%%%%%%%%%%%%%
%
We will start to discuss the case of magnetic reconnection and DSA assuming
that protons are injected into the acceleration process, as it has been
the common assumption in previous works. As the material close to the BH
originates from the helium-rich wind of the WR companion star, we 
discuss after that briefly the case of helium as CR primaries.

%%%%%%%%%%%%%%%%%%%%%%%%%%%%%%%%%%%%%%%%%%%%%%%%%%%
\begin{figure*}[t]
\includegraphics[width=\columnwidth,angle=0]{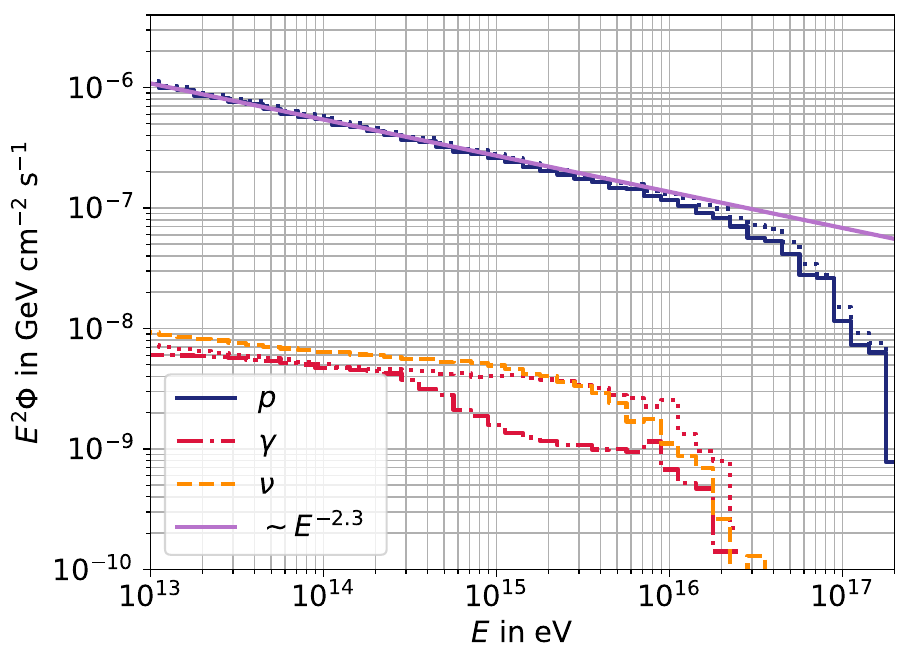}
\includegraphics[width=\columnwidth,angle=0]{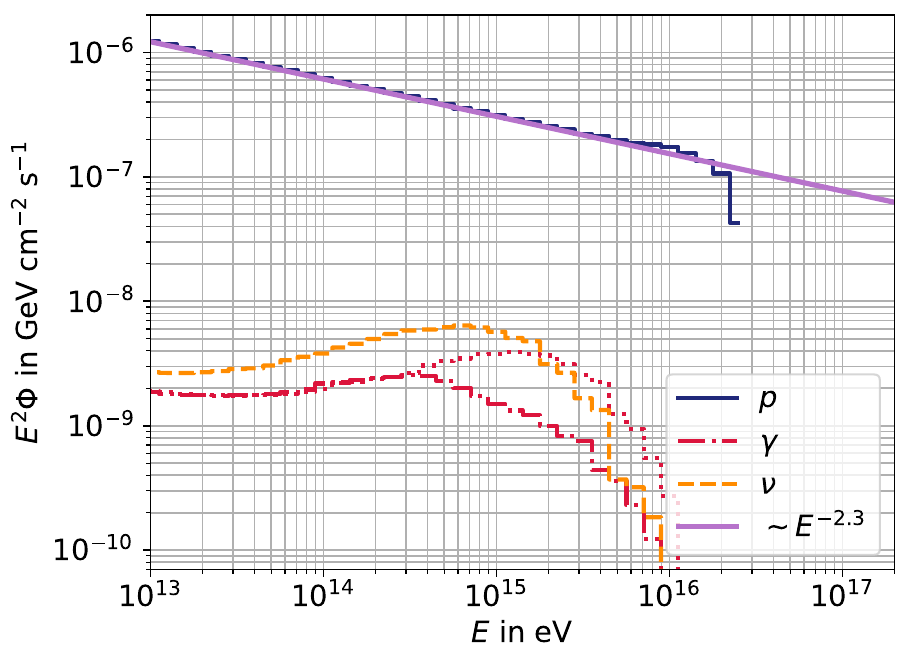}
\caption{Particle fluxes of CR protons, sum of all neutrino flavours and photon
  fluxes as function of energy $E$ for the case of reconnection
  in the state $S_1$ (left) and DSA  in the state $S_2$ (right).}
\label{fig:CR}
\end{figure*}
%%%%%%%%%%%%%%%%%%%%%%%%%%%%%%%%%%%%%%%%%%%%%%%%%%%

%%%%%%%%%%%%%%%%%%%%%%%%%%%%%%%%%%%%%%%%%%%%%%%%%%%%%%%%%%%%%%%%%%%%%%%%%%%%%  
\subsection{Proton injection}

Having specified the acceleration mechanisms and the physical parameters
determining interaction and energy loss processes, we can compare the
resulting acceleration, interaction and energy loss rates. This will allow
us to check and to interpret the numerical results from our Monte
Carlo simulations.
In Fig.~\ref{fig:rates}, we show these rates for protons\footnote{See~\citet{EL25} for a discussion of other long-lived particles.}, comparing in the
left panel the case of acceleration by reconnection in the state $S_1$
to the case of DSA  in the state $S_2$ shown on the right.
The dense photon fields close to the BH make photo-hadronic interactions
to the most important energy-loss process,
limiting the maximal energy of protons in the reconnection case to
$\simeq 6\times 10^{16}$\,eV. Hadronic interactions on gas  from the stellar
wind of the WR star play a negligible role during the acceleration phase,
even if one would assume as~\citet{Khiali:2014joa} an increased gas density
close to the BH.
In the case of DSA in the state $S_2$, synchrotron losses  and
photo-hadronic interactions are of similar importance, with a slight excess
of the former. The maximal energy of protons is somewhat smaller,
$\simeq 3\times 10^{16}$\,eV, but still high enough to expect secondaries
with energies above PeV.
Note also that in both cases the Hillas criterium allows higher CR rigidities
than the interaction and energy losses: e.g., in $S_1$, the Hillas criterium
allows for rigidites up to 300\,PV.

We normalise the fluxes such that the CR luminosity satisfies
$L_{\rm CR}=\eta_{\rm CR}\lambda L_{\rm Edd} \simeq 3.9\times 10^{38}$erg/s,
i.e.\ we assume optimistically that 15\% of the available energy is
used to accelerate hadrons.
In Fig.~\ref{fig:CR}, we show the rescaled flux
$E^2\Phi=E^2\d N/(\d A\d t \d E)$  of CR
protons (blue line), of the sum of all neutrino flavours (orange line)
and of photons (red line) as function of energy $E$. In addition to the
proton flux after interactions, the pressumed power-law $E^{-2.3}$ and the
flux escaping the acceleration region are presented.
The high-energy cutoffs agree with the expectations from Fig.~\ref{fig:rates},
i.e.\ a rather soft suppression above $8\times 10^{16}$\, eV in the state
$S_1$ and a sharper suppression above $2\times 10^{16}$\, eV in the state
$S_2$. Similarly, the secondary fluxes in the state $S_1$ are as expected
larger than in the state $S_2$.
The red dotted line indicates the unabsorbed photon flux, while the
dot-dashed line shows the photon flux taking into account pair production:
At energies above $10^{14}$\,eV
pair production on CMB photons  dominates, while at lower energies
the absorption on stellar photons becomes important.

\begin{figure*}[t]
\includegraphics[width=\columnwidth,angle=0]{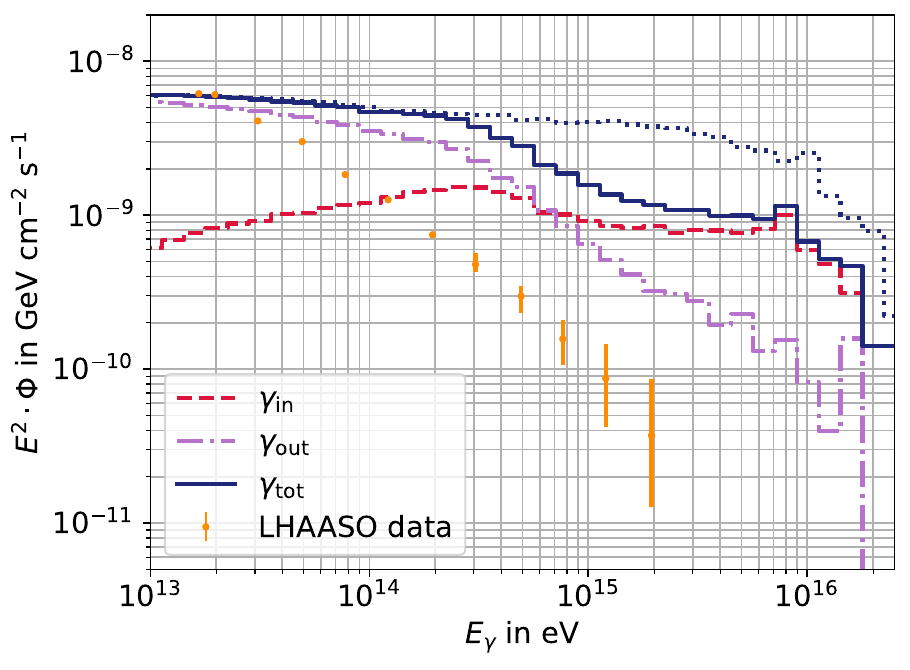}
\includegraphics[width=\columnwidth,angle=0]{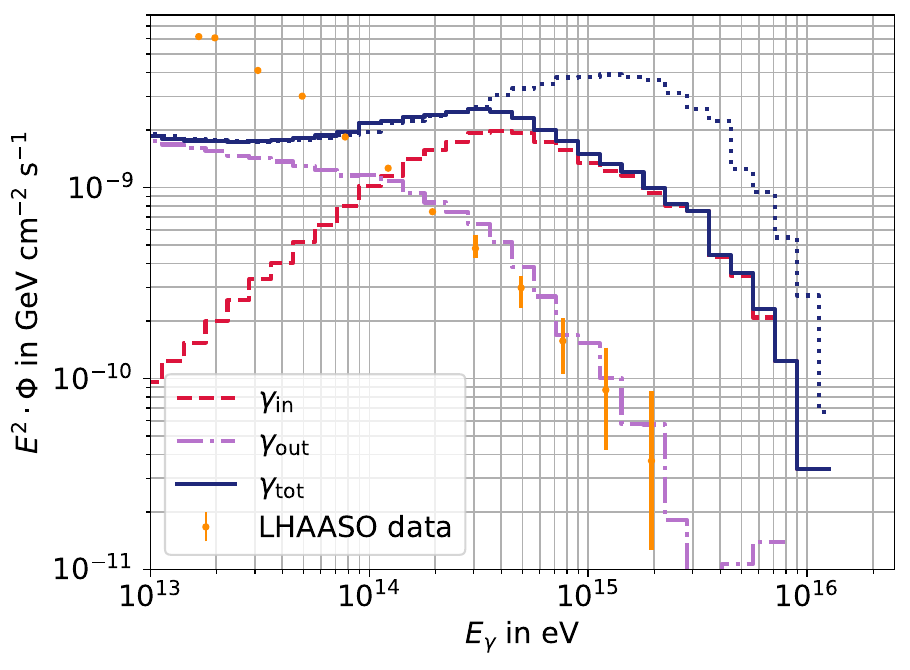}
\caption{Photon fluxes as function of energy $E$
  for the case of reconnection
  in the state $S_1$ (left) and by DSA in the state $S_2$ (right)
  compared to the  gamma-ray flux measured by
    LHAASO~\citep{LHAASO:2023uhj} in the direction of the Cygnus~X
    superbubble.}
\label{fig:gamma}
\end{figure*}

In Fig.~\ref{fig:gamma}, we show a close-up of the photon fluxes, where
we split the flux into two components, depending on their production
inside or outside the acceleration zone. The outer component is mainly
produced in hadronic interactions, and thus the slope of this component
repeats the slope of the proton flux at an energy a factor $\simeq 20$ higher.
In addition, we show in Fig.~\ref{fig:gamma} the gamma-ray flux measured
by LHAASO in the direction of the Cygnus~X superbubble
from~\citet{LHAASO:2023uhj}. Note that \C\ contributes only
a fraction, mostly at the highest energies, of the total flux from the
Cygnus superbubble. Since no detailed information on the
exposure is published, we can only roughly estimate the integrated
flux $\Phi(>E)$ from the central region above PeV energies as
$\Phi(>{\rm PeV})= N/(AT)\simeq 9\times 10^{-14}/({\rm m^2 s})$ using as
effective area $A\simeq 10^6$m$^2$ from~\citet{Ma_2022},
$T\simeq 6000$\,h for 3\,years of data taking, and $N=2$ events above PeV.
This corresponds to a photon flux of order
$E^2\Phi(E={\rm PeV})\simeq 10^{-11}$GeV/cm$^2$s.
Thus the obtained photon flux, which is in both states on the level
$E^2\Phi(E={\rm PeV})\simeq 10^{-9}$GeV/cm$^2$s,
should be scaled down by $10^{-2}$, what would reduce the
required CR acceleration efficiency down to $\eta_{\rm CR}\sim 10^{-3}$.
At this level, the photon flux would be well below the lower limits
from the MAGIC collaboration~\citep{2010ApJ...721..843A}.
The correspondingly down-scaled  neutrino flux is a factor $100$ below the
90\% C.L. upper limit set by the IceCube collaboration~\citep{IceCube:2022jpz}.
Even for $\eta_{\rm CR}= 0.1$, the number of expected muon neutrino
events/year in IceCube above 10\,TeV is only around one in $S_1$ and two
in $S_2$.

%%%%%%%%%%%%%%%%%%%%%%%%%%%%%%%%%%%%%%%%%%%%%%%%%%%
%%%%%%%%%%%%%%%%%%%%%%%%%%%%%%%%%%%%%%%%%%%%%%%%%%%

%%%%%%%%%%%%%%%%%%%%%%%%%%%%%%%%%%%%%%%%%%%%%%%%%%%%%%%%%%%%%%%%%%%%%%%%%%%%
\subsection{Helium injection}

Next we  discuss briefly the case of helium as CR primaries.
Since the acceleration and diffusion of CRs depends only on rigidity,
the rates given in Eqs.~(\ref{acc1}) and (\ref{acc2}) apply directly to
helium. As synchrotron losses for the same primary energy scale
with $(q/m)^4$, the losses of helium are reduced by a factor~16. 
In the case of hadronic interactions, the He-He cross section
is a factor 6--7 larger than the proton-proton cross section, while
the number of targets is reduced by a factor of four for a fixed wind
density $\rho$. Thus the secondary production due to hadronic interactions
is increased by a factor 1.5--1.75 relative to the proton case discussed
before.
The differences are more pronounced in the case of interactions with
background photons. Here, in addition to the usual photo-hadronic
channel, the photo dissociation of helium nuclei into nucleons sets in
at a $\simeq 10$~times lower threshold energy. Thus helium nuclei at the
highest energies will fragment into nucleons, resulting into the same energy
cutoff as obtained in the proton case. At lower energies, mainly
helium nuclei will escape from the acceleration zone, leading to a
somewhat increased secondary production. Since the interaction depth is
small, the largest part of the accelerated helium nuclei escapes and
contributes to the helium part of the Galactic CR spectrum.

%%%%%%%%%%%%%%%%%%%%%%%%%%%%%%%%%%%%%%%%%%%%%%%%%%%%
\section{Discussion}
%%%%%%%%%%%%%%%%%%%%%%%%%%%%%%%%%%%%%%%%%%%%%%%%%%%%%
%
Let us now briefly review some of our main assumptions and discuss
how our results change if we modify them. 
\begin{itemize}
\item   
Our values for the masses of the binary stars are at the higher end of the
range discussed in the literature, and the choice $M_{\rm BH}=5M_\odot$ and
$M_{\rm WR}=20M_\odot$
would be more in line with the analysis of, e.g., \citet{Koljonen:2017gvy}.
This value for $M_{\rm BH}$ would raise the temperature of the accretion disk 
by only $\sim  30\%$, and we have seen that the disk photons
play only a minor role as target. More importantly, the number density of
X-ray photons in the state $S_1$ would increase by a factor
$\simeq 4^3= 64$,
reducing the maximal energy in the $S_1$ state by a factor of order ten.
Such a reduction of the maximal proton energy would mean that \C\ is barely
a PeV photon source  in the $S_1$ state. Alternatively,
a two-zone model would have to be considered where the X-ray emission
is partly decoupled from the acceleration of hadrons.
\item % 
  The accretion rate $\dot M$ regulates the total amount of power available,
  of which the fraction $\eta_{\rm CR}\lambda$ can be channeled into the
  acceleration of CRs. X-ray observations favour super-Eddington accretion
  rates and/or a rather massive BH~\citep{Veledina:2023zho}.  Moreover,
  radio observations show a very large variability in the radio
  flux~\citep{1972Natur.239..440G,Green:2025ecw}. If this variability is
  connected to variations in 
  $\dot M$ or $\eta_{\rm CR}$, then the PeV photon flux should be in an
  one-zone model correlated with the variation in the radio band. If, on
  the other hand, a smaller jet-opening
  angle and larger jet velocities are characteristic for
  flares~\citep{10.1093/mnras/stac666}, then the PeV photon flux could be
  in the state $S_2$ anti-correlated to the radio flares.
\item   %
  In the state $S_2$, a combined scenario with both DSA at the recolliminaton
  or termination shock and 2.nd order Fermi throughout the jet or magnetic
  reconnection close the the BH could be
  possible. In particular, for a smaller value of $\delta$, the maximal
  energy achievable in DSA is reduced and this region would not contribute
  to high-energy photons. However, this region might be still the main
  contributor to the  X-ray emission via leptonic processes. Hence such a
  case would be a realisation of a two-zone model, where
  the target density for photo-hadronic interactions is strongly
  reduced, and rather small BH masses would be unproblematic even for
  the acceleration via reconnection close to the BH.
\item % 
  We have negelected relativistic beaming effects. As the flux scales with
  the Doppler factor $\delta$ as $\delta^3$, already modest gamma factors
  could lead to a strong increase of the observed photon fluxes. 
\item % 
  We have considered a purely hadronic model for the photon emission of
  \C, without trying to explain self-consistently the X-ray emission by
  electrons. A combined model where the acceleration of electrons
  and hadrons is simulated based on the same mechanism
  would help to constrain
  physical parameters like the magnetic field strength and the size of the
  acceleration region, but is deferred to a future work.
\end{itemize}

%%%%%%%%%%%%%%%%%%%%%%%%%%%%%%%%%%%%%%%%%%%%%%%%%%%%
\section{Conclusions}
%%%%%%%%%%%%%%%%%%%%%%%%%%%%%%%%%%%%%%%%%%%%%%%%%%%%%
%
We have examined the acceleration and interactions of high-energy CRs
in the high-mass X-ray binary \C, motivated by the recent observations
of two PeV photons in the direction of \C . We have found that in all
the three acceleration scenarios considered---magnetic reconnection. 2.nd
order Fermi acceleration on magnetic turbulence and diffusive shock
acceleration---CRs can be accelerated beyond PeV energies. In the hadronic
scenario studied by us, high-energy photons are mainly produced in
interactions with gas from the wind of the WR companion star. The CR
luminosity to explain the excess flux observed by LHAASO at energies
above PeV energies is rather low, requiring a CR acceleration
efficiency of $10^{-3}$--$10^{-2}$.
This suggest that the PeV photon flux from \C\  could be in a bright
phase significantly increased relative to the average flux of the
last years.

%%%%%%%%%%%%%%%%%%%%%%%%%%%%%%%%%%%%%%%%%%%%%%%%%%%%%%%%%%%%%%
\begin{acknowledgements}
We would like to thank Karri Koljonen  for useful discussions and
helpful comments on the draft, and
Egor Podlesnyi for help in estimating the expected number of IceCube events.
M.K.\ is grateful to the late Venya Berezinsky for
introducing him to the story of \C\ in the 1980s. 
\end{acknowledgements}


%\bibliographystyle{aa} 
%\bibliography{cr,cyg}

\begin{thebibliography}{51}
\expandafter\ifx\csname natexlab\endcsname\relax\def\natexlab#1{#1}\fi

\bibitem[{Abbasi {et~al.}(2022)}]{IceCube:2022jpz}
Abbasi, R. {et~al.} 2022, Astrophys. J. Lett., 930, L24

\bibitem[{{Aleksi{\'c}} {et~al.}(2010){Aleksi{\'c}}, {Antonelli}, {Antoranz},
  {Backes}, {Baixeras}, {Barrio}, {Bastieri}, {Becerra Gonz{\'a}lez},
  {Bednarek}, {Berdyugin}, {Berger}, {Bernardini}, {Biland}, {Blanch}, {Bock},
  {Boller}, {Bonnoli}, {Bordas}, {Borla Tridon}, {Bosch-Ramon}, {Bose},
  {Braun}, {Bretz}, {Britzger}, {Camara}, {Carmona}, {Carosi}, {Colin},
  {Contreras}, {Cortina}, {Costado}, {Covino}, {Dazzi}, {De Angelis}, {De Cea
  del Pozo}, {De Lotto}, {De Maria}, {De Sabata}, {Delgado Mendez}, {Doert},
  {Dom{\'\i}nguez}, {Dominis Prester}, {Dorner}, {Doro}, {Elsaesser},
  {Errando}, {Ferenc}, {Fonseca}, {Font}, {Garc{\'\i}a L{\'o}pez},
  {Garczarczyk}, {Gaug}, {Godinovic}, {G{\"o}ebel}, {Hadasch}, {Herrero},
  {Hildebrand}, {H{\"o}hne-M{\"o}nch}, {Hose}, {Hrupec}, {Hsu}, {Jogler},
  {Klepser}, {Kr{\"a}henb{\"u}hl}, {Kranich}, {La Barbera}, {Laille},
  {Leonardo}, {Lindfors}, {Lombardi}, {Longo}, {L{\'o}pez}, {Lorenz},
  {Majumdar}, {Maneva}, {Mankuzhiyil}, {Mannheim}, {Maraschi}, {Mariotti},
  {Mart{\'\i}nez}, {Mazin}, {Meucci}, {Miranda}, {Mirzoyan}, {Miyamoto},
  {Mold{\'o}n}, {Moles}, {Moralejo}, {Nieto}, {Nilsson}, {Ninkovic}, {Orito},
  {Oya}, {Paiano}, {Paoletti}, {Paredes}, {Partini}, {Pasanen}, {Pascoli},
  {Pauss}, {Pegna}, {Perez-Torres}, {Persic}, {Peruzzo}, {Prada}, {Prandini},
  {Puchades}, {Puljak}, {Reichardt}, {Rhode}, {Rib{\'o}}, {Rico}, {Rissi},
  {R{\"u}gamer}, {Saggion}, {Saito}, {Saito}, {Salvati}, {S{\'a}nchez-Conde},
  {Satalecka}, {Scalzotto}, {Scapin}, {Schultz}, {Schweizer}, {Shayduk},
  {Shore}, {Sierpowska-Bartosik}, {Sillanp{\"a}{\"a}}, {Sitarek}, {Sobczynska},
  {Spanier}, {Spiro}, {Stamerra}, {Steinke}, {Struebig}, {Suric}, {Takalo},
  {Tavecchio}, {Temnikov}, {Terzic}, {Tescaro}, {Teshima}, {Torres}, {Vankov},
  {Wagner}, {Weitzel}, {Zabalza}, {Zandanel}, {Zanin}, \& {MAGIC
  Collaboration}}]{2010ApJ...721..843A}
{Aleksi{\'c}}, J., {Antonelli}, L.~A., {Antoranz}, P., {et~al.} 2010, \apj,
  721, 843

\bibitem[{Baerwald \& Guetta(2013)}]{Baerwald:2012yd}
Baerwald, P. \& Guetta, D. 2013, Astrophys. J., 773, 159

\bibitem[{Berezinsky {et~al.}(1985)Berezinsky, Bugaev, \&
  Zaslavskaya}]{Berezinsky:1985zja}
Berezinsky, V.~S., Bugaev, E.~V., \& Zaslavskaya, E.~S. 1985, JETP Lett., 42,
  528

\bibitem[{{Berezinsky} {et~al.}(1986){Berezinsky}, {Castagnoli}, \&
  {Galeotti}}]{1986ApJ...301..235B}
{Berezinsky}, V.~S., {Castagnoli}, C., \& {Galeotti}, P. 1986, \apj, 301, 235

\bibitem[{Berezinsky {et~al.}(1986)Berezinsky, Ellis, \&
  Ioffe}]{Berezinsky:1985hr}
Berezinsky, V.~S., Ellis, J.~R., \& Ioffe, B.~L. 1986, Phys. Lett. B, 172, 423

\bibitem[{{Bhat} {et~al.}(1986){Bhat}, {Sapru}, \&
  {Razdan}}]{1986ApJ...306..587B}
{Bhat}, C.~L., {Sapru}, M.~L., \& {Razdan}, H. 1986, \apj, 306, 587

\bibitem[{Cangemi {et~al.}(2021)Cangemi, Rodriguez, Grinberg, Belmont, Laurent,
  \& Wilms}]{Cangemi:2020jaj}
Cangemi, F., Rodriguez, J., Grinberg, V., {et~al.} 2021, Astron. Astrophys.,
  645, A60

\bibitem[{Cao {et~al.}(2024)}]{LHAASO:2023uhj}
Cao, Z. {et~al.} 2024, Sci. Bull., 69, 449

\bibitem[{{Cerutti} {et~al.}(2011){Cerutti}, {Dubus}, {Malzac}, {Szostek},
  {Belmont}, {Zdziarski}, \& {Henri}}]{2011A&A...529A.120C}
{Cerutti}, B., {Dubus}, G., {Malzac}, J., {et~al.} 2011, \aap, 529, A120

\bibitem[{Chakrabarti(1996)}]{Chakrabarti:1996cc}
Chakrabarti, S.~K. 1996, Phys. Rept., 266, 229

\bibitem[{Comisso {et~al.}(2024)Comisso, Farrar, \& Muzio}]{Comisso:2024ymy}
Comisso, L., Farrar, G.~R., \& Muzio, M.~S. 2024, Astrophys. J. Lett., 977, L18

\bibitem[{Comisso \& Sironi(2019)}]{Comisso:2019frj}
Comisso, L. \& Sironi, L. 2019, Astrophys. J., 886, 122

\bibitem[{de~Gouveia Dal~Pino \& Lazarian(2005)}]{deGouveiaDalPino:2003mu}
de~Gouveia Dal~Pino, E.~M. \& Lazarian, A. 2005, Astron. Astrophys., 441, 845

\bibitem[{Fletcher {et~al.}(1994)Fletcher, Gaisser, Lipari, \&
  Stanev}]{Fletcher:1994bd}
Fletcher, R.~S., Gaisser, T.~K., Lipari, P., \& Stanev, T. 1994, Phys. Rev. D,
  50, 5710

\bibitem[{Franceschini {et~al.}(2008)Franceschini, Rodighiero, \&
  Vaccari}]{Franceschini:2008tp}
Franceschini, A., Rodighiero, G., \& Vaccari, M. 2008, Astron. Astrophys., 487,
  837

\bibitem[{Gaisser \& Stanev(1985)}]{Gaisser:1983cj}
Gaisser, T.~K. \& Stanev, T. 1985, Phys. Rev. Lett., 54, 2265

\bibitem[{Green {et~al.}(2025)Green, Rhodes, \& Bright}]{Green:2025ecw}
Green, D.~A., Rhodes, L., \& Bright, J. 2025 [\eprint[arXiv]{2502.20409}]

\bibitem[{{Gregory} {et~al.}(1972){Gregory}, {Kronberg}, {Seaquist}, {Hughes},
  {Woodsworth}, {Viner}, \& {Retallack}}]{1972Natur.239..440G}
{Gregory}, P.~C., {Kronberg}, P.~P., {Seaquist}, E.~R., {et~al.} 1972, \nat,
  239, 440

\bibitem[{Hanson {et~al.}(2000)Hanson, Still, \& Fender}]{Hanson:2000rg}
Hanson, M.~M., Still, M.~D., \& Fender, R.~P. 2000, Astrophys. J., 541, 308

\bibitem[{Hjalmarsdotter {et~al.}(2008)Hjalmarsdotter, Zdziarski, Larsson,
  Beckmann, McCollough, Hannikainen, \& Vilhu}]{Hjalmarsdotter:2007bx}
Hjalmarsdotter, L., Zdziarski, A.~A., Larsson, S., {et~al.} 2008, Mon. Not.
  Roy. Astron. Soc., 384, 278

\bibitem[{Kachelrie{\ss} {et~al.}(2008)Kachelrie{\ss}, Ostapchenko, \&
  Tom\`as}]{Kachelriess:2007tr}
Kachelrie{\ss}, M., Ostapchenko, S., \& Tom\`as, R. 2008, Phys. Rev. D, 77,
  023007

\bibitem[{Khiali {et~al.}(2015)Khiali, de~Gouveia Dal~Pino, \& del
  Valle}]{Khiali:2014joa}
Khiali, B., de~Gouveia Dal~Pino, E.~M., \& del Valle, M.~V. 2015, Mon. Not.
  Roy. Astron. Soc., 449, 34

\bibitem[{Koljonen {et~al.}(2018)Koljonen, Maccarone, McCollough, Gurwell,
  Trushkin, Pooley, Piano, \& Tavani}]{Koljonen:2017gah}
Koljonen, K. I.~I., Maccarone, T., McCollough, M.~L., {et~al.} 2018, Astron.
  Astrophys., 612, A27

\bibitem[{Koljonen \& Maccarone(2017)}]{Koljonen:2017gvy}
Koljonen, K. I.~I. \& Maccarone, T.~J. 2017, Mon. Not. Roy. Astron. Soc., 472,
  2181

\bibitem[{Koljonen {et~al.}(2023)Koljonen, Satalecka, Lindfors, \&
  Liodakis}]{Koljonen:2023xfn}
Koljonen, K. I.~I., Satalecka, K., Lindfors, E.~J., \& Liodakis, I. 2023, Mon.
  Not. Roy. Astron. Soc., 524, L89

\bibitem[{Kowal {et~al.}(2012)Kowal, de~Gouveia Dal~Pino, \& Lazarian}]{Kow12}
Kowal, G., de~Gouveia Dal~Pino, E.~M., \& Lazarian, A. 2012, Phys. Rev. Lett.,
  108, 241102

\bibitem[{Lagage \& Cesarsky(1983)}]{Lagage:1983zz}
Lagage, P.~O. \& Cesarsky, C.~J. 1983, Astron. Astrophys., 125, 249

\bibitem[{Lammert(2025)}]{EL25}
Lammert, E. 2025, Master's thesis, {TU M\"unchen}

\bibitem[{Ma {et~al.}(2022)Ma, Bi, Cao, Chen, Chen, Cheng, Gong, Gu, He, Hou,
  Huang, Huang, Liu, Shchegolev, S~heng, Stenkin, Wu, Wu, Wu, Xiao, Yao, Zhang,
  Zhang, \& Zuo}]{Ma_2022}
Ma, X.-H., Bi, Y.-J., Cao, Z., {et~al.} 2022, Chinese Physics C, 46, 030001

\bibitem[{Miller-Jones {et~al.}(2004)Miller-Jones, Blundell, Rupen,
  Mioduszewski, Duffy, \& Beasley}]{Miller-Jones:2003lop}
Miller-Jones, J. C.~A., Blundell, K.~M., Rupen, M.~P., {et~al.} 2004,
  Astrophys. J., 600, 368

\bibitem[{Mioduszewski {et~al.}(2001)Mioduszewski, Rupen, Hjellming, Pooley, \&
  Waltman}]{Mioduszewski:2001ev}
Mioduszewski, A.~J., Rupen, M.~P., Hjellming, R.~M., Pooley, G.~G., \& Waltman,
  E.~B. 2001, Astrophys. J., 553, 766

\bibitem[{Mucke {et~al.}(2000)Mucke, Engel, Rachen, Protheroe, \&
  Stanev}]{Mucke:1999yb}
Mucke, A., Engel, R., Rachen, J.~P., Protheroe, R.~J., \& Stanev, T. 2000,
  Comput. Phys. Commun., 124, 290

\bibitem[{Ostapchenko(2011)}]{Ostapchenko:2010vb}
Ostapchenko, S. 2011, Phys. Rev., D83, 014018

\bibitem[{Ostapchenko(2013)}]{Ostapchenko:2013pia}
Ostapchenko, S. 2013, EPJ Web Conf., 52, 02001

\bibitem[{{Piano} {et~al.}(2012){Piano}, {Tavani}, {Vittorini}, {Trois},
  {Giuliani}, {Bulgarelli}, {Evangelista}, {Coppi}, {Del Monte}, {Sabatini},
  {Striani}, {Donnarumma}, {Hannikainen}, {Koljonen}, {McCollough}, {Pooley},
  {Trushkin}, {Zanin}, {Barbiellini}, {Cardillo}, {Cattaneo}, {Chen},
  {Colafrancesco}, {Feroci}, {Fuschino}, {Giusti}, {Longo}, {Morselli},
  {Pellizzoni}, {Pittori}, {Pucella}, {Rapisarda}, {Rappoldi}, {Soffitta},
  {Trifoglio}, {Vercellone}, \& {Verrecchia}}]{2012A&A...545A.110P}
{Piano}, G., {Tavani}, M., {Vittorini}, V., {et~al.} 2012, \aap, 545, A110

\bibitem[{{Protheroe}(1994)}]{1994ApJS...90..883P}
{Protheroe}, R.~J. 1994, \apjs, 90, 883

\bibitem[{Reid \& Miller-Jones(2023)}]{Reid:2023ksq}
Reid, M.~J. \& Miller-Jones, J. C.~A. 2023, Astrophys. J., 959, 85

\bibitem[{Romero {et~al.}(2003)Romero, Torres, Bernado, \&
  Mirabel}]{Romero:2003td}
Romero, G.~E., Torres, D.~F., Bernado, M. M.~K., \& Mirabel, I.~F. 2003,
  Astron. Astrophys., 410, L1

\bibitem[{Sahakyan {et~al.}(2014)Sahakyan, Piano, \& Tavani}]{Sahakyan:2013opa}
Sahakyan, N., Piano, G., \& Tavani, M. 2014, Astrophys. J., 780, 29

\bibitem[{{Samorski} \& {Stamm}(1983)}]{1983ApJ...268L..17S}
{Samorski}, M. \& {Stamm}, W. 1983, \apjl, 268, L17

\bibitem[{{Schmutz} {et~al.}(1996){Schmutz}, {Geballe}, \&
  {Schild}}]{1996A&A...311L..25S}
{Schmutz}, W., {Geballe}, T.~R., \& {Schild}, H. 1996, \aap, 311, L25

\bibitem[{Shakura \& Sunyaev(1973)}]{Shakura:1972te}
Shakura, N.~I. \& Sunyaev, R.~A. 1973, Astron. Astrophys., 24, 337

\bibitem[{Spencer {et~al.}(2022)Spencer, Garrett, Bray, \&
  Green}]{10.1093/mnras/stac666}
Spencer, R.~E., Garrett, M., Bray, J.~D., \& Green, D.~A. 2022, Mon. Not. Roy.
  Astron. Soc., 512, 2618

\bibitem[{Stark \& Saia(2003)}]{Stark:2003vr}
Stark, M.~J. \& Saia, M. 2003, Astrophys. J. Lett., 587, L101

\bibitem[{Szostek \& Zdziarski(2008)}]{Szostek:2007ke}
Szostek, A. \& Zdziarski, A.~A. 2008, Mon. Not. Roy. Astron. Soc., 386, 593

\bibitem[{{van Kerkwijk} {et~al.}(1992){van Kerkwijk}, {Charles}, {Geballe},
  {King}, {Miley}, {Molnar}, {van den Heuvel}, {van der Klis}, \& {van
  Paradijs}}]{1992Natur.355..703V}
{van Kerkwijk}, M.~H., {Charles}, P.~A., {Geballe}, T.~R., {et~al.} 1992, \nat,
  355, 703

\bibitem[{Veledina {et~al.}(2024)}]{Veledina:2023zho}
Veledina, A. {et~al.} 2024, Nature Astron., 8, 1031

\bibitem[{Vernetto \& Lipari(2016)}]{Vernetto2016}
Vernetto, S. \& Lipari, P. 2016, Phys. Rev. D, 94, 063009

\bibitem[{Vilhu {et~al.}(2021)Vilhu, Kallman, Koljonen, \&
  Hannikainen}]{Vilhu:2021vfs}
Vilhu, O., Kallman, T.~R., Koljonen, K.~I., \& Hannikainen, D.~C. 2021, Astron.
  Astrophys., 649, A176

\bibitem[{{Waltman} {et~al.}(1996){Waltman}, {Foster}, {Pooley}, {Fender}, \&
  {Ghigo}}]{1996AJ....112.2690W}
{Waltman}, E.~B., {Foster}, R.~S., {Pooley}, G.~G., {Fender}, R.~P., \&
  {Ghigo}, F.~D. 1996, \aj, 112, 2690

\end{thebibliography}

\end{document}